\documentclass[a4paper,10pt,eqno]{article}
\usepackage{theorem}
\usepackage{latexsym,amssymb,amsfonts,amsmath}
\usepackage{cite}
\usepackage{color}
\newtheorem{thm}{Theorem}[section]

\newtheorem{cor}[thm]{Corollary}

\theoremheaderfont{\scshape}
\newcommand{\RM}{\mathbb{R}}
\newcommand{\ZM}{\mathbb{Z}}

\newcommand{\CM}{\mathbb{C}}

\newcommand{\ket}[1]{|#1\rangle}

\title{{\Large {\bf Asymptotic analysis of the one-dimensional quantum walks by\\ the Tsallis and R\'enyi entropies}}}
\author{
{\small Yusuke Ide\footnote{To whom correspondence should be addressed. E-mail: ide@kanagawa-u.ac.jp}}\\
{\scriptsize Department of Information Systems Creation, 
Faculty of Engineering, 
Kanagawa University}\\
{\scriptsize Kanagawa, Yokohama 221-8686, Japan}\\
{\scriptsize e-mail: ide@kanagawa-u.ac.jp}\\
{\small Norio Konno}\\
{\scriptsize Department of Applied Mathematics, 
Faculty of Engineering, 
Yokohama National University}\\
{\scriptsize Hodogaya, Yokohama 240-8501, Japan}\\
{\scriptsize e-mail: konno@ynu.ac.jp}\\
{\small Junji Shikata}\\
{\scriptsize Graduate School of Environment and Information Sciences, 
Yokohama National University}\\
{\scriptsize Hodogaya, Yokohama 240-8501, Japan}\\
{\scriptsize e-mail: shikata@ynu.ac.jp}\\
}
\date{\empty }
\begin{document}
\maketitle

\par\noindent
\begin{small}
\par\noindent
{\bf Abstract}. The Tsallis and R\'enyi entropies are important quantities in the information theory, statistics and related fields because the Tsallis entropy is an one parameter generalization of the Shannon entropy and the R\'enyi entropy includes  several useful entropy measures such as the Shannon entropy, Min-entropy and so on, as special choices of its parameter. On the other hand, the discrete-time quantum walk plays important roles in various applications, for example, quantum speed-up algorithm and universal computation. In this paper, we show limiting behaviors of the Tsallis and R\'enyi entropies for discrete-time quantum walks on the line which are starting from the origin and defined by arbitrary coin and initial state. The results show that the Tsallis entropy behaves in polynomial order of time with  the parameter dependent exponent while the R\'enyi entropy tends to infinity in logarithmic order of time independent of the choice of the parameter. Moreover, we show the difference between the R\'enyi entropy and the logarithmic function characterizes by the R\'enyi entropy of the limit distribution of the quantum walk. In addition, we show an example of asymptotic behavior of the conditional R\'enyi entropies of the quantum walk. 

\footnote[0]{
{\it Abbr. title:}
Tsallis and R\'enyi entropies for quantum walks
}
\end{small}

\setcounter{equation}{0}

\section{Introduction}
The Tsallis entropy \cite{Tsallis1988} and the R\'enyi  entropy \cite{Renyi1961} are important quantities in the information theory, statistics and related fields. The Tsallis entropy is viewed as an one parameter generalization of the Shannon entropy. Also the R\'enyi entropy includes several useful entropy measures such as the Shannon entropy, Min-entropy and so on, as special cases. In the information theory, information measures such as entropies play a significant role because they measure the quantity of information \cite{IwamotoShikata2013}. In addition, the (relative) Shannon entropy gives a meaning of the rate function of the Large Deviation Principle (LDP) for random walks.

On the other hand, the discrete-time quantum walks (DTQWs) have been attractive research topics in the last decade as quantum counterparts of the random walks \cite{Kempe2003,Kendon2007,VAndraca2012,Konno2008b,AharonovEtAl2001,AmbainisEtAl2001, ManouchehriWang2013}.  As the random walk plays important roles in various fields, DTQW also plays such roles in various applications, for example, quantum speed-up algorithm \cite{Ambainis2003,ChildsEtAl2003,Ambainis2004,ShenviEtAl2003} and universal quantum computation \cite{Childs2009,LovettEtAl2010}. Recently, the LDP for the one-dimensional DTQW have been shown in \cite{SunadaTate2012}. In the present stage, the entropic meaning of the rate function for the LDP for DTQW is unknown. Therefore it is important to make clear the roles of entropies in DTQWs.

In this paper, we investigate asymptotic behaviors of the Tsallis and R\'enyi entropies of DTQWs on the line as a fundamental study. The result for the R\'enyi entropy is a generalization of the result for the Shannon entropy given by \cite{IdeEtAl2011}. Therefore it is consistent for the numerical results of \cite{BrackenEtAl2004, ChandrashekarEtAl2008} for the Shannon entropy. 

The rest of this paper is organized as follows. In Sect. 2, we give the definition of the DTQW and state our results for the Tsallis and R\'enyi entropies (Theorem {\rmfamily \ref{thm:renyi}}) and the conditional R\'enyi entropies (Corollary {\rmfamily \ref{cor:condrenyi}}). The proof of the Theorem {\rmfamily \ref{thm:renyi}} is presented in Sect. 3. In this paper, we use two for the base of logarithm functions but the choice of the base is not essential.

\section{Definition and results}
The discrete-time quantum walk is a quantum counterpart of the classical random walk which has additional degree of freedom called chirality. For DTQW on the line $\ZM$ where $\ZM$ is the set of integers, the chirality takes two values left and right, and it means the direction of the motion of the walker. Now we define the following two dimensional vectors:
\begin{eqnarray*}
\ket{L} = 
\left[
\begin{array}{cc}
1 \\
0  
\end{array}
\right],
\qquad
\ket{R} = 
\left[
\begin{array}{cc}
0 \\
1  
\end{array}
\right],
\end{eqnarray*}
where $L$ and $R$ refer to the left and right chirality state, respectively.  
At each time step, the walker moves one step to the left if it has the left chirality, and if it has the right chirality, it moves one step to the right. 

Let $\hbox{U}(2)$ denote the set of $2 \times 2$ unitary matrices. The time evolution of the DTQW on $\ZM$ is defined by 
\begin{eqnarray*}
U =
\left[
\begin{array}{cc}
a & b \\
c & d
\end{array}
\right] \in \hbox{U}(2),
\end{eqnarray*}
with $a, b, c, d \in \CM$ where $\CM$ is the set of complex numbers. The unitarity of $U$ gives 
\begin{eqnarray*}
|a|^2 + |b|^2 =|c|^2 + |d|^2 =1, \> a \overline{c} + b \overline{d}=0, \> c= - \triangle \overline{b}, \> d= \triangle \overline{a},
\label{konno-eqn:seisitu}
\end{eqnarray*}
where $\overline{z}$ is the complex conjugate of $z \in \CM$ and $\triangle = \det U = a d - b c$ with $|\triangle|=1.$ 
In order to define the dynamics of the model, we divide $U$ into two matrices:
\begin{eqnarray*}
P =
\left[
\begin{array}{cc}
a & b \\
0 & 0 
\end{array}
\right], 
\quad
Q =
\left[
\begin{array}{cc}
0 & 0 \\
c & d 
\end{array}
\right],
\end{eqnarray*}
with $U=P+Q$. The matrix $P$ (resp. $Q$) represents the weight of the walker's movement to the left (resp. right) at each time step. 
Let $\Xi_{n}(l, m)$ denote the sum of all paths starting from the origin in the trajectory consisting of $l$ steps left and $m$ steps right. In fact, for time $n = l+m$ and position $x=-l + m$, we have 
\begin{align*}
\Xi_n (l,m) = \sum_{l_j, m_j} P^{l_{1}} Q^{m_{1}} P^{l_{2}} Q^{m_{2}} \cdots P^{l_{n-1}} Q^{m_{n-1}} P^{l_{n}} Q^{m_{n}},
\end{align*}
where the summation is taken over all integers $l_j, m_j \ge 0$ satisfying $l_1+ \cdots +l_n=l, \> m_1+ \cdots + m_n = m, \> l_j+ m_j=1$. We should note that the definition gives 
\begin{align*}
\Xi_{n+1}(l, m) = P \> \Xi_{n}(l-1, m) + Q \> \Xi_{n}(l, m-1).
\end{align*}
For example, in the case of $l=3, \> m=1$, we have 
\begin{align*}
\Xi_4 (3,1) &= QP^3 + PQP^2 + P^2QP + P^3 Q. 
\end{align*}

The set of initial qubit states at the origin for the DTQW is given by 
\begin{eqnarray}\label{eq:initq}
\Phi = \left\{ \varphi =
\alpha \ket{L}+\beta \ket{R} \in \mathbb C^2 :
|\alpha|^2 + |\beta|^2 =1
\right\}.
\end{eqnarray}
The probability that a quantum walker is in position $x$ at time $n$ starting from the origin with $\varphi \in \Phi$ is defined by 
\begin{align*}
P (X_{n}^{\varphi } =x) = || \Xi_{n}(l, m) \varphi ||^2,
\end{align*}
where $n=l+m$ and $x=-l+m$.

The Tsallis entropy $T_{\hat{\alpha }}(X)$ and the R\'enyi entropy $R_{\hat{\alpha }}(X)$ of order $\hat{\alpha }\in [0,\infty )\setminus \{1\}$ for a random variable $X$ taking values in a finite set $\mathcal{X}$ are given by
\begin{align*}
T_{\hat{\alpha }}(X) &= \frac{1}{1-\hat{\alpha }} \left(  \sum_{x\in \mathcal{X}} \> P_{X}(x)^{\hat{\alpha }}-1 \right),\\
R_{\hat{\alpha }}(X) &= \frac{1}{1-\hat{\alpha }} \log_2  \left\{ \sum_{x\in \mathcal{X}} \> P_{X}(x)^{\hat{\alpha }} \right\},
\end{align*}
where $P_{X}$ is a probability measure on $\mathcal{X}$. Let $P_n^{\varphi } (x) = P (X_{n}^{\varphi } =x)$. Here we define the Tsallis entropy of order $\hat{\alpha }\in [0,\infty )\setminus \{1\}$ for the DTQW at time $n$ as follows:
\begin{align}\label{eq:defTsallis}
T_{\hat{\alpha }}^{\varphi }(n) =  T_{\hat{\alpha }}(X_{n}^{\varphi }) = \frac{1}{1-\hat{\alpha }} \left( \sum_{x=-n}^n \> P_n^{\varphi } (x)^{\hat{\alpha }}-1 \right).
\end{align}
Also we define the R\'enyi entropy of order $\hat{\alpha }\in [0,\infty )\setminus \{1\}$ for the DTQW at time $n$ as follows:
\begin{align}\label{eq:defRenyi}
R_{\hat{\alpha }}^{\varphi }(n) =  R_{\hat{\alpha }}(X_{n}^{\varphi }) = \frac{1}{1-\hat{\alpha }} \log_2  \left\{ \sum_{x=-n}^n \> P_n^{\varphi } (x)^{\hat{\alpha }} \right\}.
\end{align}
Note that the limit $\lim _{\hat{\alpha }\to 1}R_{\hat{\alpha }}^{\varphi }(n)$ is the Shannon entropy $S_n^{\varphi } = - \sum_{x=-n}^n \> P_n^{\varphi } (x) \log_2 P_n^{\varphi } (x)$. Also $R_{\infty }^{\varphi }:=\lim _{\hat{\alpha }\to \infty}R_{\hat{\alpha }}^{\varphi }(n)$ is equal to $-\log \max _{-n\leq x \leq n}P_n^{\varphi } (x)$ the min-entropy. Remark that the Tsallis entropy and the R\'enyi entropy defined by Eqs. (\ref{eq:defTsallis}) and (\ref{eq:defRenyi}) have the following one-to-one correspondence:
\begin{align*}
T_{\hat{\alpha }}^{\varphi }(n) = \frac{1}{1-\hat{\alpha }} \left( 2^{(1-\hat{\alpha })R_{\hat{\alpha }}^{\varphi }(n) }-1 \right).
\end{align*}
In this paper, we show the following long-time behavior of the Tsallis entropy and the R\'enyi entropy:

\begin{thm}
\label{thm:renyi}
If the DTQW is determined by $U$ with $abcd \not= 0$, then we have
\begin{align*} 
\lim_{n\to \infty }\left\{\frac{T_{\hat{\alpha }}^{\varphi }(n) + 1/(1-\hat{\alpha })}{(n/2)^{1-\hat{\alpha }}}-\frac{1}{1-\hat{\alpha }}\right\}
=
\frac{1}{1-\hat{\alpha }}\left( \int_{-|a|}^{|a|} f(x)^{\hat{\alpha }} dx -1 \right),
\end{align*}
and 
\begin{align} 
\lim_{n \to \infty} \log_2 (n/2) \left( \frac{R_{\hat{\alpha }}^{\varphi }(n)}{\log_2 (n/2)} -1 \right) 
&= \frac{1}{1-\hat{\alpha }} \log_2  \left\{ \int_{-|a|}^{|a|} f(x)^{\hat{\alpha }} dx \right\}
=:R_{\hat{\alpha }}^{\varphi }(\infty ),
\label{manj2}
\end{align}
for each $\hat{\alpha } \in [0,\infty )\setminus \{1\}$. Here 
\begin{align*}
f(x) = \frac{|b|}{\pi (1-x^2) \sqrt{|a|^2 - x^2}} \left\{   1 - \left( |\alpha|^2 - |\beta|^2 + \frac{ a \alpha \overline{b \beta} + \overline{a \alpha} b \beta }{|a|^2} \right) x \right\}.
\end{align*}
\end{thm}
Theorem \ref{thm:renyi} shows that the asymptotic behavior of the Tsallis entropy $T_{\hat{\alpha }}^{\varphi }(n)$ is governed by $(n/2)^{1-\hat{\alpha }}$ while the R\'enyi entropy $R_{\hat{\alpha }}^{\varphi }(n)$ tends to infinity in order $\log _2(n/2)$ independent of choice of the parameter $\hat{\alpha }$. Moreover, the difference between the R\'enyi entropy $R_{\hat{\alpha }}^{\varphi }(n)$ and $\log _2(n/2)$ is measured by the R\'enyi entropy with related to $f(x)$ the limit density function of $X_{n}^{\varphi }/n$ which is obtained by \cite{Konno2002, Konno2005}. This situation is consistent with the Shannon entropy case \cite{IdeEtAl2011}. It is observed that asymptotic behaviors of the Tsallis and R\'enyi entropies are strongly dependent on that of $X_{n}^{\varphi }/n$ in any choice of the parameter $\hat{\alpha }$. 

It is natural to consider the conditional entropy for DTQW for the next step. In this paper, we consider the conditional R\'enyi entropies. As it is already mentioned by \cite{IwamotoShikata2013}, form the axiomatic point of view, there are five types of definitions of the conditional R\'enyi entropies of order $\hat{\alpha }\in [0,\infty )\setminus \{1\}$ for the DTQW at time $n$ as follows:
\begin{align*}
R_{\hat{\alpha }}^{C}(X_{n}^{\varphi }|Y) &=
\sum_{y\in \mathcal{Y}}P_{Y}(y)R_{\hat{\alpha }}(X_{n}^{\varphi }|Y=y),
\\
R_{\hat{\alpha }}^{JA}(X_{n}^{\varphi }|Y) &=
R_{\hat{\alpha }}(X_{n}^{\varphi },Y) - R_{\hat{\alpha }}(Y),
\\
R_{\hat{\alpha }}^{RW}(X_{n}^{\varphi }|Y) &=
\frac{1}{1-\hat{\alpha }}\max_{y\in \mathcal{Y}} \left[ \log_2  \left\{ \sum_{x=-n}^n \> P_{X_{n}^{\varphi }|Y} (x|y)^{\hat{\alpha }} \right\}\right],
\\
R_{\hat{\alpha }}^{A}(X_{n}^{\varphi }|Y) &=
\frac{\hat{\alpha }}{1-\hat{\alpha }} \log_2 \left[ \sum_{y\in \mathcal{Y}}P_{Y}(y)\left\{ \sum_{x=-n}^n \> P_{X_{n}^{\varphi }|Y} (x|y)^{\hat{\alpha }} \right\}^{1/\hat{\alpha }}\right],
\\
R_{\hat{\alpha }}^{H}(X_{n}^{\varphi }|Y) &=
\frac{1}{1-\hat{\alpha }} \log_2  \left[ \sum_{y\in \mathcal{Y}}P_{Y}(y) \left\{ \sum_{x=-n}^n \> P_{X_{n}^{\varphi }|Y} (x|y)^{\hat{\alpha }} \right\}\right],
\end{align*}
where $Y$ is a random variable taking values in a finite set $\mathcal{Y}$. 

Now we examine the case that the initial qubit state $\varphi $ of DTQW is randomly chosen by a random variable $Y$ with some probability measure $P_{Y}$ on $\mathcal{Y}\subset \Phi $ defined by Eq. (\ref{eq:initq}). In this setting, the conditional probability measure $P_{X_{n}^{\varphi }|Y}$ is the same as $P_n^{\varphi }$ on the condition $\{Y=\varphi \}$. In addition, $ \log_2  \left\{ \sum_{x=-n}^n \> P_n^{\varphi } (x)^{\hat{\alpha }} \right\}=(1-\hat{\alpha })R_{\hat{\alpha }}^{\varphi }(n)$ by the definition of the R\'enyi entropy Eq. (\ref{eq:defRenyi}). Then these conditional entropies are calculated by the following forms:
\begin{align*}
R_{\hat{\alpha }}^{C}(X_{n}^{\varphi }|Y) &=
\sum_{\varphi \in \mathcal{Y}}P_{Y}(\varphi )R_{\hat{\alpha }}^{\varphi }(n),
\\
R_{\hat{\alpha }}^{JA}(X_{n}^{\varphi }|Y) &=
\frac{1}{1-\hat{\alpha }} \log_2  \left \{ \sum_{\varphi \in \mathcal{Y}}P_{Y}(\varphi )^{\hat{\alpha }}\cdot 2^{(1-\hat{\alpha })R_{\hat{\alpha }}^{\varphi }(n)} \right \}
 - R_{\hat{\alpha }}(Y),
\\
R_{\hat{\alpha }}^{RW}(X_{n}^{\varphi }|Y) &=
\begin{cases}
\displaystyle\max_{\varphi \in \mathcal{Y}}\left \{R_{\hat{\alpha }}^{\varphi }(n) \right \}, &\text{if $0\leq \hat{\alpha } <1$,}\\
\displaystyle\min_{\varphi \in \mathcal{Y}}\left \{R_{\hat{\alpha }}^{\varphi }(n) \right \}, &\text{if $1 < \hat{\alpha }$,}
\end{cases}
\\
R_{\hat{\alpha }}^{A}(X_{n}^{\varphi }|Y) &=
\frac{\hat{\alpha }}{1-\hat{\alpha }} \log_2  \left \{ \sum_{\varphi \in \mathcal{Y}}P_{Y}(\varphi )\cdot 2^{\left((1-\hat{\alpha })/\hat{\alpha } \right)R_{\hat{\alpha }}^{\varphi }(n)} \right \},
\\
R_{\hat{\alpha }}^{H}(X_{n}^{\varphi }|Y) &=
\frac{1}{1-\hat{\alpha }} \log_2  \left \{ \sum_{\varphi \in \mathcal{Y}}P_{Y}(\varphi )\cdot 2^{(1-\hat{\alpha })R_{\hat{\alpha }}^{\varphi }(n)} \right \}.
\end{align*}
As a consequence, we have the following corollary by using Theorem \ref{thm:renyi}:
\begin{cor}\label{cor:condrenyi}
If the DTQW is determined by $U$ with $abcd \not= 0$, then we have the limits of the conditional R\'enyi entropies for each $\hat{\alpha } \in [0,\infty )\setminus \{1\}$.
\begin{align*}
\lim_{n\to \infty}\log_{2}(n/2)\left(\frac{R_{\hat{\alpha }}^{C}(X_{n}^{\varphi }|Y)}{\log_{2}(n/2)}-1 \right) &=
\sum_{\varphi \in \mathcal{Y}}P_{Y}(\varphi )R_{\hat{\alpha }}^{\varphi }(\infty ),
\\
\lim_{n\to \infty}\log_{2}(n/2)\left(\frac{R_{\hat{\alpha }}^{JA}(X_{n}^{\varphi }|Y)}{\log_{2}(n/2)}-1 \right) &=
\frac{1}{1-\hat{\alpha }} \log_2  \left \{ \sum_{\varphi \in \mathcal{Y}}P_{Y}(\varphi )^{\hat{\alpha }}\cdot 2^{(1-\hat{\alpha })R_{\hat{\alpha }}^{\varphi }(\infty )} \right \}
 - R_{\hat{\alpha }}(Y),
\\
\lim_{n\to \infty}\log_{2}(n/2)\left(\frac{R_{\hat{\alpha }}^{RW}(X_{n}^{\varphi }|Y)}{\log_{2}(n/2)}-1 \right) &=
\begin{cases}
\displaystyle\max_{\varphi \in \mathcal{Y}}\left \{R_{\hat{\alpha }}^{\varphi }(\infty ) \right \}, &\text{if $0\leq \hat{\alpha } <1$,}\\
\displaystyle\min_{\varphi \in \mathcal{Y}}\left \{R_{\hat{\alpha }}^{\varphi }(\infty ) \right \}, &\text{if $1 < \hat{\alpha }$,}
\end{cases}
\\
\lim_{n\to \infty}\log_{2}(n/2)\left(\frac{R_{\hat{\alpha }}^{A}(X_{n}^{\varphi }|Y)}{\log_{2}(n/2)}-1 \right) &=
\frac{\hat{\alpha }}{1-\hat{\alpha }} \log_2  \left \{ \sum_{\varphi \in \mathcal{Y}}P_{Y}(\varphi )\cdot 2^{\left((1-\hat{\alpha })/\hat{\alpha } \right)R_{\hat{\alpha }}^{\varphi }(\infty )} \right \},
\\
\lim_{n\to \infty}\log_{2}(n/2)\left(\frac{R_{\hat{\alpha }}^{H}(X_{n}^{\varphi }|Y)}{\log_{2}(n/2)}-1 \right) &=
\frac{1}{1-\hat{\alpha }} \log_2  \left \{ \sum_{\varphi \in \mathcal{Y}}P_{Y}(\varphi )\cdot 2^{(1-\hat{\alpha })R_{\hat{\alpha }}^{\varphi }(\infty )} \right \},
\end{align*}
where $R_{\hat{\alpha }}^{\varphi }(\infty )$ is defined by Eq. (\ref{manj2}).
\end{cor}

\section{Proof of Theorem {\rmfamily \ref{thm:renyi}}}
In this section we assume $abcd \not=0$. We consider the following four matrices: 
\begin{eqnarray*}
P =
\left[
\begin{array}{cc}
a & b \\
0 & 0 
\end{array}
\right], 
\quad
Q =
\left[
\begin{array}{cc}
0 & 0 \\
c & d 
\end{array}
\right],
\quad 
R =
\left[
\begin{array}{cc}
c & d \\
0 & 0 
\end{array}
\right], 
\quad
S =
\left[
\begin{array}{cc}
0 & 0 \\
a & b 
\end{array}
\right]. 
\end{eqnarray*}
Put $x \wedge y = \min \{x,y \}$. For $l \wedge m \ge 1$, we have 
\begin{align*}
\Xi_n (l,m) 
= a^l \overline{a}^m \triangle^m \sum_{\gamma =1}^{l \wedge m} \left(- \frac{|b|^2}{|a|^2} \right)^{\gamma} {l-1 \choose \gamma-1} {m-1 \choose \gamma -1} \left( \frac{l - \gamma}{a \gamma} P + \frac{m - \gamma}{\triangle \overline{a} \gamma} Q - \frac{1}{\triangle \overline{b}} R + \frac{1}{b} S\right), 
\end{align*}
by the path counting method \cite{Konno2002, Konno2005}. Therefore, for $k \in \{1, \ldots ,[n/2]\}$, we have
\begin{align*}
P_{n}^{\varphi}(n-2k)
&
=  |a|^{2(n-1)} 
\sum_{\gamma =1} ^{k} \sum_{\delta =1} ^{k}
\left(-{|b|^2 \over |a|^2} \right)^{\gamma + \delta} 
{k-1 \choose \gamma- 1} 
{k-1 \choose \delta- 1} 
{n-k-1 \choose \gamma- 1} 
{n-k-1 \choose \delta- 1} 
\\
& 
\qquad \qquad 
\times
\left( {1 \over \gamma \delta} \right) \>
\Biggl[ 
\{ k^2 |a|^2 + (n-k)^2|b|^2 - (\gamma + \delta) (n-k)\} |\alpha|^2 
\\
&
\qquad \qquad \qquad \qquad \qquad
+
\{ k^2 |b|^2 + (n-k)^2|a|^2  - (\gamma + \delta) k \} |\beta|^2 
\\
&
\qquad \qquad \qquad \qquad \qquad 
+ {1 \over |b|^2}
\biggl[ 
\{ (n-k) \gamma - k \delta + n(2k-n) |b|^2 \} a \alpha \overline{b \beta} 
\\
&
\qquad \qquad \qquad \qquad \qquad \qquad 
+ 
\{ -k  \gamma +(n-k) \delta + n(2k-n) |b|^2 \} \overline{a \alpha} b \beta + 
\gamma \delta 
\biggr] \Biggr],
\end{align*}
\begin{align*}
P_{n}^{\varphi}(-(n-2k))
&
=  |a|^{2(n-1)} 
\sum_{\gamma =1} ^{k} \sum_{\delta =1} ^{k}
\left(-{|b|^2 \over |a|^2} \right)^{\gamma + \delta} 
{k-1 \choose \gamma- 1} 
{k-1 \choose \delta- 1} 
{n-k-1 \choose \gamma- 1} 
{n-k-1 \choose \delta- 1} 
\\
& 
\qquad \qquad 
\times
\left( {1 \over \gamma \delta} \right) \>
\Biggl[ 
\{ k^2 |b|^2 + (n-k)^2|a|^2 - (\gamma + \delta) k \} |\alpha|^2 
\\
&
\qquad \qquad \qquad \qquad \qquad
+
\{ k^2 |a|^2 + (n-k)^2|b|^2 - (\gamma + \delta) (n-k) \} |\beta|^2 
\\
&
\qquad \qquad \qquad \qquad \qquad
+ {1 \over |b|^2}
\biggl[ 
\{ k \gamma - (n-k) \delta - n(2k-n) |b|^2 \} a \alpha \overline{b \beta} 
\\
&
\qquad \qquad \qquad \qquad \qquad \qquad 
+ 
\{ -(n-k) \gamma + k \delta - n(2k-n) |b|^2 \} \overline{a \alpha} b \beta + 
\gamma \delta 
\biggr] \Biggr],
\\
P_{n}^{\varphi}(n)
&
= |a|^{2(n-1)} \{ |b|^2 |\alpha|^2 +|a|^2 |\beta|^2- (a \alpha \overline{b \beta} + \overline{a \alpha} b \beta ) \}, 
\\
P_{n}^{\varphi}(-n) 
&
= |a|^{2(n-1)} \{ |a|^2 |\alpha|^2 +|b|^2 |\beta|^2+ (a \alpha \overline{b \beta} + \overline{a \alpha} b \beta ) \}.
\end{align*} 
Here $[x]$ denote the integer part of $x \in \RM$ where $\RM$ is the set of real numbers.  

Let $P^{\nu, \mu} _n (x)$ denote the Jacobi polynomial which is orthogonal on $[-1,1]$ with respect to $(1-x)^{\nu}(1+x)^{\mu}$ with $\nu, \mu > -1$. Then the following relation holds:
\begin{align*}
P^{\nu, \mu} _n (x) = \frac{\Gamma (n + \nu + 1)}{\Gamma (n+1) \Gamma (\nu +1)} \> {}_2F_1(- n, n + \nu + \mu +1; \nu +1 ;(1-x)/2),
\end{align*}
where ${}_2F_1(a, b; c ;z)$ is the hypergeometric series and $\Gamma (z)$ is the gamma function. In general, as for orthogonal polynomials, see \cite{Andrews1999}. Then we have
\begin{align}
\sum_{\gamma =1} ^{k}
\left(- \frac{|b|^2}{|a|^2} \right)^{\gamma -1}
{1 \over \gamma} 
{k-1 \choose \gamma- 1}  
{n-k-1 \choose \gamma- 1} 
&= 
\frac{|a|^{-2(k-1)}}{k} P^{1,n-2k} _{k-1}(2|a|^2-1), 
\label{yukari1}
\\
\sum_{\gamma =1} ^{k}
\left(- \frac{|b|^2}{|a|^2} \right)^{\gamma -1}
{k-1 \choose \gamma- 1}  
{n-k-1 \choose \gamma- 1} 
&= |a|^{-2(k-1)} P^{0,n-2k} _{k-1}(2|a|^2-1).
\label{yukari2}
\end{align}
By using Eqs. (\ref{yukari1}) and (\ref{yukari2}), we see that for $k \in \{1, \ldots ,[n/2]\}$, 

\begin{align*}
P_{n}^{\varphi}(\pm (n-2k))
&=
|a|^{2n - 4k -2}|b|^4 / 2\\
&\times 
\biggl[ 
\left\{ {2x^2-2x+1 \over x^2} (P^{1})^2 - {2 \over x} P^{1} P^{0}+
{2 \over |b|^2} (P^{0})^2 \right\} \\
& \qquad \pm \left( {1-2x \over x} \right) 
\biggl\{ - {1 \over x} 
\{ 
(|a|^2 - |b|^2) 
(|\alpha|^2 - |\beta|^2) + 2 (a \alpha \overline{b \beta} + \overline{a \alpha} b \beta ) \} (P^{1})^2 \\
& \qquad \qquad \qquad \qquad \qquad \qquad \qquad \qquad -2 \left( |\alpha|^2 - |\beta|^2 - 
{ a \alpha \overline{b \beta} + \overline{a \alpha} b \beta \over |b|^2 } 
\right) P^{0} P^{1} \biggl\} 
\biggr].
\end{align*}
where $P^{i} = P^{i,n-2l} _{l-1}(2|a|^2-1) \> (i=0,1)$. 
Let $a(n) \sim b(n)$ means $a(n)/b(n) \to 1$ as $n \to \infty$. As it is mentioned in \cite{Konno2002, Konno2005}, if $n\to \infty $ with $k/n\in (-(1-|a|)/2, (1+|a|)/2)$, then 
\begin{align*}
P^{0}&\sim \frac{2|a|^{2k-n}}{\sqrt{\pi n\sqrt{-\Lambda }}}\cos(An+B), \\
P^{1}&\sim \frac{2|a|^{2k-n}}{\sqrt{\pi n\sqrt{-\Lambda }}}\sqrt{\frac{x}{(1-x)(1-|a|^2)}}\cos(An+B+\theta ),
\end{align*}
where $\Lambda =(1-|a|^2)\{(2x-1)^2-|a|^2\}$, $A$ and $B$ are some constants which are independent of $n$, and $\theta \in [0,\pi/2]$ is determined by $\cos \theta =\sqrt{(1-|a|^2)/4x(1-x)}$. By these asymptotics and the Riemann-Lebesgue lemma, we obtain
\begin{align*}
R_{\hat{\alpha }}^{\varphi}(n)
&=  \frac{1}{1-\hat{\alpha } } \log_2  \left\{ \sum_{x=-n}^n \> P_n^{\varphi} (x)^{\hat{\alpha }} \right\}
\\
&\sim \frac{1}{1-\hat{\alpha }} \log_2  \left\{ \int_{(1-|a|)/2}^{(1+|a|)/2} f(1-2x)^{\hat{\alpha }} dx \times \frac{ 1 }{ n^{ \hat{\alpha }-1 } }\right\}
\\
&= \log_2 n + \frac{1}{1-\hat{\alpha }} \log_2  \left\{ \int_{-|a|}^{|a|} f(x)^{\hat{\alpha }} dx  \times 2^{ \hat{\alpha }-1 } \right\}
\\
&= \log_2 (n/2) + \frac{1}{1-\hat{\alpha }} \log_2  \left\{ \int_{-|a|}^{|a|} f(x)^{\hat{\alpha }} dx \right\}.
\end{align*}
By the same argument, we have 
\begin{align*}
T_{\hat{\alpha }}^{\varphi}(n) + \frac{1}{1-\hat{\alpha}}
=  \frac{1}{1-\hat{\alpha } } \sum_{x=-n}^n \> P_n^{\varphi} (x)^{\hat{\alpha }}
\sim \left(\frac{n}{2}\right)^{ 1-\hat{\alpha }} \times \frac{1}{1-\hat{\alpha }} \int_{-|a|}^{|a|} f(x)^{\hat{\alpha }} dx,
\end{align*}
as $n \to \infty$. This completes the proof.

\section{Summary}
In this paper, we show a limit theorem for the Tsallis and R\'enyi entropies of the DTQWs on $\ZM$ starting from the origin with arbitrary coin and initial state by using a path counting method. The result shows that asymptotic behaviors of both entropies are strongly dependent on that of scaling limit of the walker's position in any choice of the parameter. It is an interesting future problem that building some novel information theoretic scheme by using DTQWs and these entropies. Also searching for the entropic meaning of the LDP for the DTQW is one of important future problems to be solved.

\par
\
\par\noindent
{\bf Acknowledgments.}  
This work was partially supported by the Grant-in-Aid for Scientific Research (C) of Japan Society for the Promotion of Science (Grant No. 24540116).


\begin{small}

\end{small}


\begin{thebibliography}{000}



\bibitem{AharonovEtAl2001}
Aharonov, D., Ambainis, A., Kempe, J., Vazirani, U.:
Quantum walks on graphs. 
In: Proceedings of the 33rd Annual ACM Symposium on Theory of Computing, pp. 50--59 (2001).


\bibitem{Ambainis2003}
Ambainis, A.: 
Quantum walks and their algorithmic applications. 
Int. J. Quantum Inf. {\bf 1}, 507--518 (2003).


\bibitem{Ambainis2004} 
Ambainis, A.: 
Quantum walk algorithm for element distinctness. 
In: Proceedings of the 45th Annual IEEE Symposium on Foundations of Computer Science, pp. 22--31 (2004).


\bibitem{AmbainisEtAl2001} 
Ambainis, A., Bach, E., Nayak, A., Vishwanath, A., Watrous, J.: 
One-dimensional quantum walks. 
In: Proceedings of the 33rd Annual ACM Symposium on Theory of Computing, pp. 37--49 (2001).


\bibitem{Andrews1999} 
Andrews, G. E., Askey, R., Roy, R.: 
Special Functions. Cambridge University Press (1999).


\bibitem{BrackenEtAl2004} 
Bracken, A. J., Ellinas, D., Tsohantjis, I.:  
Pseudo memory effects, majorization and entropy in quantum random walks, 
J. Phys. A : Math. Gen. {\bf 37}, L91--L97 (2004).


\bibitem{ChandrashekarEtAl2008}
Chandrashekar, C. M., Srikanth, R., Laflamme, R.: 
Optimizing the discrete time quantum walk using a $SU(2)$ coin. 
Phys. Rev. A {\bf 77}, 032326 (2008).


\bibitem{Childs2009}
Childs, A. M.:
Universal computation by quantum walk. 
Phys. Rev. Lett. {\bf 102}, 180501 (2009).


\bibitem{ChildsEtAl2003} 
Childs, A. M., Cleve, R., Deotto, E., Farhi, E., Gutmann, S., Spielman, D. A.: 
Exponential algorithmic speedup by quantum walk. 
In: Proceedings of the 35rd Annual ACM Symposium on Theory of Computing, pp. 59--68 (2003).


\bibitem{IdeEtAl2011}
Ide, Y., Konno, N., Machida, T.:
Entanglement for discrete-time quantum walks on the line.
Quant. Inf. Comput. {\bf 11}, 0855--0866 (2011).


\bibitem{IwamotoShikata2013}
Iwamoto, M., Shikata, J.:
Information Theoretic Security for Encryption Based on Conditional R\'enyi Entropies.
The 7th International Conference on Information Theoretic Security (ICITS2013), LNCS 8317, pp.103-121, Springer, November 2013 (2013). 
The full version is entitled ``Revisiting Conditional R\'enyi Entropies and Generalizing Shannon's Bounds in Information Theoretically Secure 
Encryption'', and available at http://eprint.iacr.org/2013/440 .


\bibitem{Kempe2003} 
Kempe, J.: 
Quantum random walks - an introductory overview. 
Contemporary Physics {\bf 44},  307--327 (2003).


\bibitem{Kendon2007} 
Kendon, V.: 
Decoherence in quantum walks - a review. 
Math. Struct. in Comp. Sci. {\bf 17}, 1169--1220 (2007).


\bibitem{Konno2002} 
Konno, N.: 
Quantum random walks in one dimension. 
Quant. Inform. Process {\bf 1}, 345--354 (2002).


\bibitem{Konno2005} 
Konno, N.: 
A new type of limit theorems for the one-dimensional quantum random walk. 
J. Math. Soc. Jpn. {\bf 57}, 1179--1195 (2005).


\bibitem{Konno2008b} 
Konno, N.: 
Quantum Walks. 
In: Quantum Potential Theory, Franz, U., and Sch\"urmann, M., Eds., 
Lecture Notes in Mathematics: Vol. 1954, pp. 309--452, Springer-Verlag, Heidelberg (2008).


\bibitem{LovettEtAl2010}
Lovett, N. B., Cooper, S., Everitt, M., Trevers, M., Kendon, V.: 
Universal quantum computation using the discrete-time quantum walk.
Phys. Rev. A {\bf 81}, 042330 (2010).


\bibitem{ManouchehriWang2013}
Manouchehri, K., Wang, J.B.:
Physical Implementation of Quantum Walks. 
Springer-Verlag, Heidelberg (2013).


\bibitem{Renyi1961}
R\'enyi, A.: 
On measures of entropy and information. 
In: Proceedings of the 4th Berkeley Symposium on Mathematics, Statistics and Probability 1960, pp. 547--561 (1961).


\bibitem{ShenviEtAl2003} 
Shenvi, N., Kempe, J., Whaley, K. B.: 
Quantum random-walk search algorithm. 
Phys. Rev. A {\bf 67}, 052307 (2003).


\bibitem{SunadaTate2012}
Sunada, T., Tate, T.:
Asymptotic behavior of quantum walks on the line.
J. Funct. Anal. {\bf 262}, 2608--2645 (2012).


\bibitem{Tsallis1988}
Tsallis, C.:
Possible generalization of Boltzmann-Gibbs statistics.
J. Stat. Phys. {\bf 52}, 479--487 (1988).


\bibitem{VAndraca2012}
Venegas-Andraca, S.E.
Quantum walks: A comprehensive review.
Quant. Inform. Process {\bf 11}, 1015--1106 (2012).


\end{thebibliography}
\end{document}